\documentclass[12pt]{article}
\usepackage{amsmath}
\usepackage{graphicx,psfrag,epsf}
\usepackage{enumerate}
\usepackage{algorithm}
\usepackage{algorithmic}
\usepackage{float} 
\usepackage{url} 
\usepackage{cite}
\usepackage{array} 
\usepackage{amssymb}

\usepackage{makecell}

\usepackage{gensymb}

\newcommand{\blind}{0}

\begin{document}

\def\spacingset#1{\renewcommand{\baselinestretch}%
{#1}\small\normalsize} \spacingset{1}


\if0\blind
{
  \title{\bf Numerical Optimization of Material Configurations for the Development of Surface Plasmon Resonance Devices }
  \author{Casey Kneale, Karl S. Booksh\hspace{.2cm}\\
    Department of Chemistry and Biochemistry\\
    University of Delaware, Newark, Delaware 19716\\
	}
  \maketitle
} \fi

\if1\blind
{
  \bigskip
  \bigskip
  \bigskip
  \begin{center}
    {\LARGE\bf }
\end{center}
  \medskip
} \fi

\bigskip
\begin{abstract}
Various metals (Ag, Al, Au, Bi, Cu), polymers (polyvinylpyrrolidone, polystyrene), and electrically conductive polymers (polyacetylene, polyaniline, poly(3,4-ethylenedio-xythiophene) polystyrene sulfonate) were subjected to a particle swarm optimizer in both the planar and grating configuration to optimize conditions which supported surface plasmon resonances (SPR) for chemical sensing. The objective functions for these configurations were based on absorption peak depth, full width at half maximum, or the enhancement factor (planar). Simple logic gates were constructed for both configurations which assessed if a lossy region was plasmonic by  several figures of merit. The planar substrates returned viable sensing configurations for all of the metals tested, most notably bismuth metal at ~2.1$\mu$m. The corrugated metal substrates also resulted in tuned SPRs. Most interestingly an optimized surface plasmon on the conductive polymer, polyaniline at 12$\mu$m, was also discovered. 
\end{abstract}

\noindent%

\newpage
\spacingset{1.45} 
\section{Introduction}
\label{sec:intro}
Surface plasmon resonance (SPR) sensors have become an integral technique for the selective quantitation of biomolecules, their interactions, gasses, and many other analytes. The operating principal of SPR is founded upon the photoexcitation of delocalized electrons from a host material into a resonant mode which perturbs an adjacent medium. The resonant mode, or SPR, can be completely described by the energy loss due to excitation, the material properties such as dielectric constant, and the optical coupling conditions for excitation. Sensors which operate on this principal exploit the fact that as adjacent medium changes in chemical nature, so does the excitation energy for the mode. There are a great deal of analytical schemes which have been developed that rely on this phenomena for chemical sensing. However, little work has been performed on the theoretical improvement of plasmonic hosts.

Some of the earliest works in interpolating SPR spectra to material properties were based on finding the thickness of a plasmonic material from it's dielectric constants, coupling conditions, and spectral minima \cite{Kretschmann}. For purposes of optimizing SPR sensors, the inverse of their approach is most pertinent. That is to say, from a given dielectric function of a plasmonic host can  optimal coupling conditions and material thicknesses/geometries for sensing parameters be found. Most methods which approached this problem or it's inverse utilized Lorentzian curve fitting and other less rigorous means until recently \cite{ChenChen}\cite{Fontana}.
	
Recently, Cavalcanti et al., demonstrated that the use of lorentzian fitting was not required and Monte Carlo or Partical Swarm Optimization (PSO) methods can be used to optimize SPR sensitivity when coupled with Fresnel equations. The authors left the frequency of incident radiation constant and focused on tuning gold thickness and incident angle \cite{Cavalcanti}. Their methodology is effective for tuning materials which are known to host plasmons. However, it does not account for the optimization of wavelength, or reject other modes which may present as minima in reflectance spectra. Nor is it applicable to corrugated media.

In this work, schemes for both the discovery  and optimization of surface plasmon resonances for planar and periodically corrugated materials are presented. When assessing novel hosts for plasmonics the primary concern should be the strength and clarity of a signal. The methods contained herein can also be applied in order to find experimental conditions for both Surface Enhanced Raman Spectroscopy (SERS), and SPR sensor development. However, we place emphasis on mode discrimination and novel materials due to Cavalcanti et. al's efforts toward particle swarm optimizer tuning of sensitivity.

\section{Planar Approach}
\label{sec:planar}
The planar optimization problem has several features which allow for simple exploitation to optimizer frameworks. For example, the enhancement factor (E) element of the Fresnel scattering matrix ($|S_{1,1}^2|$), has been shown to serve as a discriminator for plasmon behavior and purely absorbing modes \cite{Nature}. Similarly, the avoidance of brewster modes and other absorbing phenomena can be accomplished by a framework which we have previously described \cite{FigOfMerit}. These criteria ensure that an optimizer is selective toward the SPR phenomenon. Although unreported, the other modes illustrated in the aforementioned work can also be implemented to this framework for the optimization of brewster modes for chemical sensing.

We utilized a variation of a particle swarm optimizer, the Crystallization Particle Swarm Optimizer (CPSO), which we have previously described \cite{CPSO}. This algorithm was selected because in tests it proved to be robust to local minima and valley functions and the movement of particles can be conveniently based on experimental resolution. The independent variables available to the CPSO algorithm were wavelength, incident angle, and plasmonic host thickness. Schemes for obtaining minima and FWHM were based on an adaptive procedure.

At every iteration of the PSO algorithm the particles relocate. After relocation an angular, or frequency based interrogation is pursued with a preset window (IE: $\pm$10 nm, or $\pm$ 15\degree). Maxima for E and minima for reflection and transmission are then assessed inside of the window. FWHM's were calculated by simple arguments based on calculus and distances from an observed extrema. If a FWHM exists and the criteria for a genuine SPR mode have been met, then the objective function (O) to be minimized was assessed as,
$$O = \frac{FWHM}{E}$$ 

For general SPR material optimization applications, the enhancement factor is an obvious parameter for optimization. However, if one were interested in chemical sensing instrumentation or SERS stability then sensitivity (S) and peak depth/signal strength (P) construct a more relevant but unweighted objective function,
$$O = \frac{FWHM}{S \cdot P}$$ 
 
\section{Planar Results and Discussion}
\label{sec:planarRes}
Transition metals (Au, Ag,  Cu), polymeric materials (polyvinylpyrrolidone, polystyrene), and three metals (Al, Bi, Sb) were implemented into the optimizer scheme. These materials were used as the plasmonic host in the Kretschmann configuration with a soda lime glass substrate, and water or air semi-infinite ambient layer (Table I). The CPSO algorithm was restricted to 21 particles and 1000 iterations.

\begin{table}[H]
	\begin{center}
		\caption{Figures of merit obtained for various materials in the Kretschmann configuration. Entries denoted b, utilized an upper half space of air, and those denoted N/A did not feature figures of merit.} \label{tab:tabone}
		\resizebox{\textwidth}{!}{
		\begin{tabular}{r|cccccc}
Host &E	&FWHM ($^{\circ}$)	&Wavelength (nm)	&Thickness (nm)	&Excitation Angle ($^{\circ}$)	&\\\hline
Ag	&41.80	&0.51	&998.43	&48.03	&63.20	&\\
Al$^b$	&4.59	&0.66	&529.60	&36.25	&41.72	&\\
Au	&48.97	&0.49	&996.30	&57.00	&63.68	&\\
Bi	&104.63	&0.20	&2,092	&81.88	&61.36	&\\
Cu	&130.03	&0.17	&1,241	&53.20	&62.46	&\\
polystyrene	&N/A	&N/A	&N/A	&N/A	&N/A	&\\
polyvinylpyrrolidone	&N/A	&N/A	&N/A	&N/A	&N/A	&\\

		\end{tabular}
}
\end{center}
\end{table} 

The existence of surface plasmon resonances occurring on Au, Ag, Al, and Cu have been well documented. The optimized results were proved to be plasmonic in nature by electric field calculations. Bismuth was found to exhibit a strong plasmonic field enhancement relative to the transition metals at 2.092 $\mu$m. Although, Bismuth has been shown to exhibit plasmon resonances in grating configurations \cite{Bi} to the best of our knowledge we believe this is the first report of optimized conditions which could support SPR on planar bismuth for sensing purposes. This plasmon is especially interesting because there currently exists several commercially available photodiodes with photosensitivity maximums conveniently located around 2.1 $\mu$m. 

As previously tested, the method did not return figures of merit for the two polymeric substrates. Both of the polymer substrates were utilized as negative experimental controls for false positive SPR identification. In the visible region, polyvinylpyrrolidone, possess a relatively small extinction coefficient. Polystyrene has only a refractive index. Although, these tests offer credibility to the figures of merit employed for plasmon characterization, less ideal results were obtained for electrically conductive polymers which have proportionally large extinction coefficients.

A previous experimental investigation performed by our research group attempted to experimentally find surface plasmon resonances on the conductive polymer polyaniline \cite{PaniOne} \cite{PaniTwo} in the visible and near infrared regions of the electromagnetic spectrum. Although modes and the chemical nature of polyaniline films were found to have analytic applications, firm evidence of plasmonic behavior had not been realized. Utilizing the methodology herein, optimizations were performed on dielectric constants which were experimentally obtained for 6 different conductive polyaniline species, 2 sets for polyacetylene, and 2 for poly(3,4-ethylenedioxythiophene) (PEDOT). All of which returned inconclusive results from field calculations or brewster mode false positives for the visible region.

The false positives which occurred were believed to be a result of using too large of a step size in the optimizers' angular interrogations. This practice was necessary for computational efficiency. Although true SPRs in planar configurations were not found by this approach, it is still possible that waveguide mode coupled SPRs or purely absorbing modes could be found.

\section{Grating Approach}
\label{sec:GratingApp}
Corrugated materials offer different conditions in which light can couple to delocalized electrons of a metal. An ideal approach for solving Maxwell's equations for plane waves impinging upon a grating is the coordinate transformation method created by Jean Chandezon. The Chandezon method features small scattering matrices which leads to efficient computations, represents grating geometries in a natural manner, and allows for accurate solutions to deep gratings. For a full review of this rigorous technique, please see the following publications: \cite{OrigCMeth}, \cite{AdaptiveSpatial}, \cite{DiffFormalism}. 

The grating configuration also introduces two new optimization parameters for the CPSO algorithm. These parameters are the period of the grating, and the grating amplitude. The geometry of the grating could also be considered a free parameter. From an experimental perspective, grating geometry is typically limited by availability of lithographic techniques and the periodicity of the desired grating/wavelength of incident light. For our circumstances, sinuisoidal grating patterns are the most feasible to obtain.

The computational expense of allowing period, amplitude, layer depth, incident angle and wavelength to be free parameters was deemed inefficient. Instead, only period, amplitude, and layer depth were available to the CPSO. At each of these locations in ${\rm I\!R}^3$ a low resolution angular and wavelength interrogation was performed over a desired spectral region. The wavelength regions were selected from dispersion rules and relationships in the dielectric functions. Wherever conservation of energy minima were observed a finer resolution was implemented in an adaptive manner.

Unlike the planar case, the enhancement factor (E) is not readily obtainable. A different scheme for mode discrimination with a similar objective function was employed. Surface plasmon resonances, like many other phenomena, absorb energy. Thus by conservation of energy, for an SPR to exist there must be an absorption minima in the sum of the reflected and transmitted orders of the grating. To ensure that absorption losses were attributable to resonant modes the C-figure \cite{FigOfMerit} was employed at the locations of such extrema. The C-figure tends to return values which are orders of magnitude lower than unity if an SPR was present at a given coupling condition, and a value greater than unity if a nonresonant lossy phenomena was present.

The variables which were optimized by CPSO were the unabsorbed conservation of energy (U), and the piecewise multiplicatively weighted FWHM. We define the unabsorbed conservation of energy as,
$$U = 1-((1-B) - P)$$
Where B is the average baseline on both sides of the peak and U $\in [0, 1]$. With this figure of merit, the weighted FWHM describes the simple objective function,
$$O = \widetilde{FWHM} \cdot {U}$$

\section{Grating Results}
\label{sec:GratingRes}
Similar to the planar case, several metals (Au, Ag, Al, Cu) and polymeric materials were implemented into the optimizer schema as sinusoidal gratings (Table II). The CPSO algorithm utilized 13 particles, performed a maximum of 220 iterations per particle, and was limited to shallow grating amplitudes relative to the incident light. FWHM values found below 1.75 $^{\circ}$ were weighted via the addition of 5 times the difference from that threshold. The weighting was performed to disfavor shallower spectral bands at the potential cost of finer FWHMs.

\begin{table}[H]
	\begin{center}
		\caption{Optimized conditions and figures of merit for materials in the multilayer configuration of vacuum, plasmonic host, soda lime glass. Entries denoted N/A did not feature  a C-figure less than unity at an absorbtion peak.} \label{tab:tabtwo}
		\resizebox{\textwidth}{!}{
			\begin{tabular}{r|cccccc}
				&Height &Period &Depth 	& Wavelength & FWHM  & Absorbance \\
				Host & (nm)	& (nm)	& (nm)	&  (nm)	& ($^{\circ}$) &	($\%$) \\\hline
				Ag	& 48.17 & 1227	& 223.3	& 575	& 1.749	& 83.03\\
				Al	& 51.05 & 1162 & 487 & 775	& 1.770 & 76.82 \\
				Au	& 41.33 & 1053  	& 690 &	625 & 1.749 & 85.93\\
				Bi	& 2399 & 18811 & 2354 & 13100 & 1.741  & 9.97\\
				Cu	& 55.80  & 1695  & 189 & 700 & 1.749  & 88.15\\
				polyvinylpyrrolidone	& N/A & N/A	& N/A	& N/A	& N/A	& N/A\\
			\end{tabular}
		}   
	\end{center}
\end{table} 

Au, Ag, Al, and Cu were found to host many resonant modes which could be utilized for chemical sensing. As in the planar case, the figure of merit avoided the modeling of nonresonant but lossy features incurred by polyvinylpyrrolidone in the visible light regime. Polystyrene was not investigated because materials which lack an imaginary component to their dielectric functions present without losses in conservation of energy so long as proper convergence is attained. Conductive polymers showed more promise for hosting of SPRs in the grating configuration than in the planar case.

In the visible to near infrared range the conductive polymers tested here-in (polyacetylene, polyaniline, PEDOT:PSS) lacked features remniscent of SPRs. None of the optimized results could be utilized for chemical sensing. Several results satisfied the C-figure but they afforded such low peak  absorbances (IE: $<$ 1x$10^{-2}$) that even if they were SPR's, they would likely be nondistinguishable from experimental background. Our investigations near 1$\mu$m, a convenient location for near infrared and visible spectrometers, for finding plasmonic excitations were not fruitful. In several dielectric datasets, including one which was collected in house, the imaginary permitivity was rarely larger then the real. However, in the infrared region some success was met for conductive polyanilines (Table III).

\begin{table}[H]
	\begin{center}
		\caption{Optimized conditions and figures of merit for materials in the multilayer configuration of vacuum, polyaniline, soda lime glass. Entries denoted N/A did not feature  a C-figure less than unity at an absorbtion peak or an absorbance of $>$1$\%$. Sensitivity was calculated with an additional corrugated layer that had a refractive index 1.01 and a depth of 16$\mu$m.} \label{tab:tabthree}
		\resizebox{\textwidth}{!}{
			\begin{tabular}{r|ccccccc}
				&Height &Period &Depth 	& Wavelength & FWHM  & Absorbance & S\\
				Host & (nm)	& (nm)	& (nm)	&  (nm)	& ($^{\circ}$) &	($\%$)  & ($\frac{^{\circ}}{RIU}$)\\\hline
				
				Polyaniline	& N/A  & N/A  	& N/A	& N/A & N/A & N/A &  N/A \\
				(900-1700nm)& &	& & & & & \\
				
				Polyaniline	&1540  & 16987 & 2519& 12000& 1.766 & 20.29 & 75 \\
				(8000-12000nm)	& & & & & & & \\
				
			\end{tabular}
		} 
	\end{center}
\end{table} 

Most published dielectric functions of conductive polyanilines show that a bound surface wave can exist from approximately 1-30 $\mu$m. The strongest resonance (C-figure = 2.9x$10^{-6}$ at truncation order = 15) was found at 12. $\mu$m or 833.33 cm$^{-1}$. Which is well with-in the fourier transform infrared spectroscopy range. The plasmon at this location was about twice as absorbing as the plasmon found on thermally evaporated bismuth at 13.1$\mu$m. A similar result was demonstrated by Monas \cite{MonasThesis}, which reported a bound resonance near 10.6$\mu$m. Ultimately, the relaxed coupling conditions which were afforded by the grating configuration allowed the search for plasmons on these materials more successful than in the planar investigations. There were however, practical concerns with the grating approach we have employed.

From an experimental perspective the summation of all diffracted and transmitted modes of a grating may be impractical for an SPR or SERs design. Typically the diffraction/transmission angles for each order makes such measurements difficult without specialized mirrors or lenses to direct the light to a detector pupil. 

In these investigations, most of the optimizations returned material thicknesses where diffracted plane waves were orders of magnitude greater in relative intensity than transmissive orders. Which implies that the contribution of SPR absorbtions were typically localized to a few diffracted orders. Thus, the orders which offer the greatest figures of merit (signal strength, sensitivity, etc) could then be selected for a single diffracted order instrument design. 

If such a design was of interest the objective functions may be constructed and evaluated for the individual diffracted/transmitted orders' respective figures of merit. One advantage to the utilization of PSO algorithms for this variety of investigation was that there were thirteen particle best optima which could be investigated for these desired properties. Such that, an individual assessment of the orders did not appear beneficial. Regardless,  that level of interpretation was not necessary for our purposes.

\section{Conclusion}
\label{sec:Conclusion}
Schemes for the optimization of surface plasmon resonances occurring on planar and grating configurations were introduced and evaluated. The planar approach was able to discriminate between surface plasmons and other lossy modes for most materials using efficient run time characteristics. A promising plasmon was found to be excited on planar bismuth metal at 2.092$\mu$m. 

The corrugated approach also discovered optimized conditions for surface plasmon resonances on a variety of materials. Although most of the SPRs which were optimized were found on metals, this optimization scheme was able to optimize plasmons on conducting polyanilines in 3 layer materials.

\section{Data for Materials}
\label{sec:Origin}
The information pertaining to dielectric constants, or refractive indices, were obtained for the various materials via several references. Linear interpolations were performed between data points from the available data to allow for finer resolutions at the cost of approximation.

\begin{table}[H]
	\begin{center}
		\resizebox{\textwidth}{!}{
			\begin{tabular}{r|l}
				Material & Reference \\\hline
				Ag	& H.-J. Hagemann, W. Gudat, and C. Kunz.  J. Opt. Soc. Am. 65, 742-744 (1975)\\
				
				Al	& A. D. Raki\'c, A. B. Djurisic, J. M. Elazar, and M. L. Majewski. Appl. Opt. 37, 5271-5283 (1998)\\
				
				Au	& Meye M.; Etchegoin, P. G.; Le Ru, E. C. ; The Journal of Chemical Physics. 2006, 125, 164705.\\
				
				Bi	& Khalilzadeh-Rezaie, F.; Smith, C. W.; Nath, J.; Nader, N.; Shahzad, M.; Cleary, J. W.;\\
				&  Avrutsky, I.; Peale, R. E.Journal of Nanophotonics 2015, 9 (1), 093792. \\
				
				Cu	&  A. D. Raki\'c, A. B. Djurisic, J. M. Elazar, and M. L. Majewski. Appl. Opt. 37, 5271-5283 (1998) \\
				
				Soda Lime Glass & M. Rubin. Solar Energy Materials 12, 275-288 (1985)\\
				
				PEDOT:PSS & J. Gasiorowski et al.  Thin Solid Films. 536 (2013), 211–215\\
				(600-1400nm) & \\
				
				Polyacetylene & G. Leising.  Synthetic Metals, 28,  D215-D223 (1989)\\
				(350-900nm) & \\
				
				Polyaniline  &Tzamalis, Georgios (2002) Optical and transport properties of polyaniline films,Durham theses,\\
				(8000-12000nm) &  Durham
				University. Available at Durham E-Theses Online: http://etheses.dur.ac.uk/4148/ \\
				
				Polystyrene & N. Sultanova, S. Kasarova and I. Nikolov. Acta Physica Polonica A 116, 585-587 (2009)\\
				
				Polyvinylpyrrolidone & T. A. F. K\"onig; P. A. Ledin; J. Kerszulis; M. A. Mahmoud; M. A. El-Sayed;  \\
				&  J. R. Reynolds; V. V. Tsukruk. ACS Nano 8, 6182-6192 (2014)\\
				
				Water & G. M. Hale and M. R. Querry. Appl. Opt. 12, 555-563 (1973) \\
				
			\end{tabular}
		}
	\end{center}
\end{table}

\end{document}